\documentclass[%
 aip,
 amsmath,amssymb,
 reprint,%
]{revtex4-1}

\usepackage{graphicx}
\usepackage{dcolumn}
\usepackage{circuitikz}
    \usetikzlibrary{arrows}
\usepackage{bm}
\usepackage{url}

\usepackage[utf8]{inputenc}
\usepackage[T1]{fontenc}
\usepackage{mathptmx}
\usepackage{etoolbox}
\usepackage[font = bf]{subcaption}
\usepackage{caption}
\makeatletter
\def\@email#1#2{%
 \endgroup
 \patchcmd{\titleblock@produce}
  {\frontmatter@RRAPformat}
  {\frontmatter@RRAPformat{\produce@RRAP{*#1\href{mailto:#2}{#2}}}\frontmatter@RRAPformat}
  {}{}
}%
\makeatother

\begin{document}

\title{Frequency-domain multiplexing of SNSPDs with tunable superconducting resonators}

\author{Sasha Sypkens}
\affiliation{Arizona State University, Tempe, AZ 85281 USA}
\affiliation{Jet Propulsion Laboratory, California Institute of Technology, 4800 Oak Grove Dr., Pasadena, CA 91109, USA}

\author{Lorenzo Minutolo}
\affiliation{California Institute of Technology, 1200 E California Blvd., Pasadena, CA 91125, USA}

\author{Sahil Patel}
\affiliation{California Institute of Technology, 1200 E California Blvd., Pasadena, CA 91125, USA}
\affiliation{Jet Propulsion Laboratory, California Institute of Technology, 4800 Oak Grove Dr., Pasadena, CA 91109, USA}

\author{Emanuel Knehr}
\affiliation{Jet Propulsion Laboratory, California Institute of Technology, 4800 Oak Grove Dr., Pasadena, CA 91109, USA}

\author{Alexander B. Walter}
\affiliation{Jet Propulsion Laboratory, California Institute of Technology, 4800 Oak Grove Dr., Pasadena, CA 91109, USA}

\author{Henry G. Leduc}
\affiliation{Jet Propulsion Laboratory, California Institute of Technology, 4800 Oak Grove Dr., Pasadena, CA 91109, USA}

\author{Lautaro Narváez}
\affiliation{California Institute of Technology, 1200 E California Blvd., Pasadena, CA 91125, USA}

\author{Ralph Chamberlin}
\affiliation{Arizona State University, Tempe, AZ 85281 USA}

\author{Tracee Jamison-Hooks}
\affiliation{Arizona State University, Tempe, AZ 85281 USA}

\author{Matthew D. Shaw}
\affiliation{Jet Propulsion Laboratory, California Institute of Technology, 4800 Oak Grove Dr., Pasadena, CA 91109, USA}

\author{Peter K. Day}
\affiliation{Jet Propulsion Laboratory, California Institute of Technology, 4800 Oak Grove Dr., Pasadena, CA 91109, USA}

\author{Boris Korzh}
\affiliation{Jet Propulsion Laboratory, California Institute of Technology, 4800 Oak Grove Dr., Pasadena, CA 91109, USA}

\begin{abstract}
This work culminates in a demonstration of an alternative Frequency Domain Multiplexing (FDM) scheme for Superconducting Nanowire Single-Photon Detectors (SNSPDs) using the Kinetic inductance Parametric UP-converter (KPUP) made out of NbTiN.  There are multiple multiplexing architectures for SNSPDs that are already in use, but FDM could prove superior in applications where the operational bias currents are very low, especially for mid- and far-infrared SNSPDs. Previous FDM schemes integrated the SNSPD within the resonator, while in this work we use an external resonator, which gives more flexibility to optimize the SNSPD architecture. The KPUP is a DC-biased superconducting resonator in which a nanowire is used as its inductive element to enable sensitivity to current perturbations.  When coupled to an SNSPD, the KPUP can be used to read out current pulses on the few $\mu$A scale.  The KPUP is made out of NbTiN, which has high non-linear kinetic inductance for increased sensitivity at higher current bias and high operating temperature.  Meanwhile, the SNSPD is made from WSi, which is a popular material for broadband SNSPDs. To read out the KPUP and SNSPD array, a software-defined radio platform and a graphics processing unit are used. Frequency Domain Multiplexed SNSPDs have applications in astronomy, remote sensing, exoplanet science, dark matter detection, and quantum sensing.
\end{abstract}


\maketitle 

\section{Introduction}

Superconducting Nanowire Single-Photon Detectors (SNSPDs) have been in development for many years~\cite{Goltsman2001} and have primarily dominated in the visible/near-infrared wavelengths in terms of efficiency~\cite{Marsili2013, Reddy2020, Chang2021, Los2024}, dark count rates~\cite{Chiles2022}, count rates~\cite{Craiciu23, Resta2023}, timing resolution~\cite{Korzh2020}, and array formats~\cite{Oripov2023}.  Recently, SNSPD groups have been pushing the long-wavelength cut-off to the mid-infrared (mid-IR)~\cite{Verma2021, Chen2021, Pan2022, Taylor2022} and most recently the far-infrared (far-IR, >25~$\mu$m)~\cite{Taylor2023}.  SNSPDs that detect longer wavelength photons tend to have lower operational bias currents, which means their pulse heights are lower~\cite{gregortaylor2021}. Although there are already many multiplexing schemes for SNSPDs~\cite{Zhao2017, Wollman2019, Allmaras2020, Kong2022, McCaughan2022, Hampel2023}, Frequency Domain Multiplexing (FDM) may have an advantage in Mid-IR detection due to their high sensitivity to current perturbations. 

Frequency Domain Multiplexed (FDM) SNSPDs have been demonstrated up to 16 pixels \cite{Knehr, Doerner_2017}. In these demonstrations, an SNSPD is embedded within a microwave resonator as the inductive element, with a capacitive element in parallel.  These RF-SNSPDs are biased from a RF probe tone that is coupled from the transmission line and through the nanowire section of the resonator.  This removes the need for individual bias lines for each SNSPD, but does mean that the bias current is oscillating, which reduces the duty cycle of the maximum detection efficiency.  Another technique used for FDM SNSPDs is to still use the SNSPD as the inductive element of microwave resonator, but to DC bias the nanowires instead of solely relying on the RF probe tone~\cite{Sinclair_2019}.

This paper demonstrates a proof-of-concept for a different approach to FDM SNSPDs by using an external current-sensitive resonator to read out the SNSPDs. This method separates the components such that there is a compact photon collecting area for the SNSPDs and a readout area for the readout resonators.  The resonator in this configuration is a highly sensitive superconducting current sensor known as the Kinetic inductance Parametric UP-converter (KPUP)~\cite{Kher2016}.

\section{Experimental Setup}

\subsection{Operating Principle}

\begin{figure*}
\centering
\ctikzset{bipoles/thickness=1}
\begin{circuitikz}[scale=0.58,line width=0.7pt]

    \draw
    (-16,-5) to node[ground]{} (-16,-5)
    (-16,-5) to [vco, /tikz/circuitikz/bipoles/length=1.1cm] (-16,-3)
    (-16,-3) to [resistor=$Z_0$, /tikz/circuitikz/bipoles/length=0.8cm] (-16,-1)
    (-16,-1) -- (-14.5,-1)
    (-14,-2) to [resistor, /tikz/circuitikz/bipoles/length=0.8cm] (-14,0)
    (-13.5,-1) -- (-11.5,-1)
    (-11,-2) to [resistor, /tikz/circuitikz/bipoles/length=0.8cm] (-11,0)
    (-10.5,-1) -- (-3,-1) to [TL] (-1,-1) -- (1,-1) to [TL] (3,-1) -- (6,-1) 
    (6,-1) to [amp, /tikz/circuitikz/bipoles/length=1.1cm] (7.5,-1) -- (11,-1) -- (11,-2)
    (11,-2) to [resistor=$Z_L$, /tikz/circuitikz/bipoles/length=0.8cm] (11,-4)
    (11,-4) to node[ground]{} (11,-4)

    (0,-1) to [capacitor = $C_c$, /tikz/circuitikz/bipoles/length=0.8cm] (0,-3)
    (-2,-4.5) to [capacitor = $C_r$, /tikz/circuitikz/bipoles/length=0.8cm] (2,-4.5)
    (-2,-3) to [variable american inductor, /tikz/circuitikz/bipoles/length=0.8cm] (0,-3)
    (0,-3) to [variable american inductor, /tikz/circuitikz/bipoles/length=0.8cm] (2,-3)
    (-2,-3) -- (-2,-5.75)
    (2,-3) -- (2,-5.75)
    (-2,-5.9) -- (-1,-5.9) -- (-1,-8)
    (-2,-5.75) to [capacitor, /tikz/circuitikz/bipoles/length=0.8cm] (-2,-7.25) 
    (-2,-6.6) to node[ground]{} (-2,-6.75)
    (2,-5.75) to [capacitor, /tikz/circuitikz/bipoles/length=0.8cm] (2,-7.25) 
    (2,-6.6) to node[ground]{} (2,-6.75)
    (0,-8.5) node[text width=1.7cm, align = center]{KPUP\:chip}

    (-4.75,-8) -- (-8.25,-8) 
    (-10.25,-8) to [resistor = 500$\Omega$, /tikz/circuitikz/bipoles/length=0.8cm] (-8.25,-8) 
    (-10.25,-8) -- (-13,-8) -- (-13,-7.55)
    (-2,-5.5) node[crossingshape, scale = 2]{}
    (-4.75,-4) -- (-8.25,-4) 
    (-10.25,-4) to [resistor = 500$\Omega$, /tikz/circuitikz/bipoles/length=0.8cm] (-8.25,-4)  
    (-10.25,-4) -- (-13,-4)
    (-3.25, -8) -- (1.25, -8)
    (-4.75,-8) to [resistor = 50$\Omega$, /tikz/circuitikz/bipoles/length=0.8cm] (-3.25,-8)
    (-3.25,-5.5) -- (-2.15,-5.5)
    (-4.75,-5.5) to [resistor = 50$\Omega$, /tikz/circuitikz/bipoles/length=0.8cm] (-3.25,-5.5)

    (-13,-7) to node[vsourceAMshape, /tikz/circuitikz/bipoles/length=1.1cm, rotate = -90]{} (-13,-7)
    (-13,-6.45) to [resistor, /tikz/circuitikz/bipoles/length=0.8cm] (-13,-4)
    (-4.75,-4) -- (-4.75,-5.5)
    (-14.5,-7) node[text width=1cm, align = center]{SNSPD bias}
    
    (-7.25,-4) to [variable resistor, /tikz/circuitikz/bipoles/length=0.8cm] (-7.25,-6)
    (-6.75,-8) to [american inductor, /tikz/circuitikz/bipoles/length=0.8cm] (-6.75,-6)
    (-6.75,-8.5) node[text width=1.7cm, align = center]{SNSPD\:chip}
    (-6.25,-6) to [switch, /tikz/circuitikz/bipoles/length=0.8cm] (-6.25,-4)
    (-7.25,-6) -- (-6.25,-6)

    (10, -6.45) -- (10,-5.5) to [resistor, /tikz/circuitikz/bipoles/length=0.8cm] (7.5,-5.5)
    (10,-7) to node[vsourceAMshape, /tikz/circuitikz/bipoles/length=1.1cm, rotate = -90]{} (10,-7)
    (10,-7.55) -- (10,-8) -- (7.5,-8)
    (1,-8) -- (3.25,-8) to [american inductor = 0.8nH, /tikz/circuitikz/bipoles/length=0.8cm] (5.75,-8) -- (7.5,-8)
    (3.25,-5.5) to [american inductor = 0.8nH, /tikz/circuitikz/bipoles/length=0.8cm] (5.75,-5.5) -- (7.5,-5.5)
    (3.25,-5.5) -- (-1.85,-5.5)
    (11.5,-7) node[text width=1cm, align = center]{KPUP bias};

    \draw
    (-7,-11) rectangle (0.75,-15)
    (-4.5,-12.5) rectangle (-6.75,-14.5)
    (-5.25,-13.5) node[text width=1.5cm]{Xilinx Kintex-7\\FPGA}
    (-3.25,-10.5) node[text width=1.5cm, align = center]{SDR}
    (-4.25,-12.5) -- (-2.75,-12.5) -- (-2.5,-12.9) -- (-2.75,-13.3) -- (-4.25,-13.3) -- (-4.25,-12.5)
    (-3.6,-12.9) node[text width=1.5cm, align = center]{DAC}
    (-4.25,-13.7) -- (-2.75,-13.7) -- (-2.5,-14.1) -- (-2.75,-14.5) -- (-4.25,-14.5) -- (-4.25,-13.7)
    (-3.6,-14.1) node[text width=1.5cm, align = center]{ADC}
    (-2.3,-11.25) rectangle (0.5,-14.8)
    (-0.95,-11.75) node[text width=1.5cm, align = center]{RF board}
    (-0.95,-12.9) node[text width=1.5cm, align = center]{Tx}
    (-0.95,-14.1) node[text width=1.5cm, align = center]{Rx}
    (-2.05,-12.9) -- (-1.5,-12.9)
    (-0.25,-12.9) -- (0.25,-12.9)
    (-2.05,-14.1) -- (-1.5,-14.1)
    (-0.25,-14.1) -- (0.25,-14.1)

    ;\draw[black] (2.25,-12.5) rectangle (4.25,-14.5);
    \fill[black!50!green, opacity = 0.1] (2.25,-12.5) rectangle (4.25,-14.5);\draw
    (3.25,-13.5) node[align = center]{KPUP}
    (0.75,-12.9) -- (2.25,-12.9)
    (0.75,-14.1) -- (2.25,-14.1)
    (-10.5,-12.25) rectangle (-8.5,-13.75)
    (-9.5,-13) node[text width=1.7cm, align = center]{GPU}
    
    (-10.25,-15.5) -- (4,-15.5)
    (-3.25,-16) node[text width=3cm, align = center]{Digital Readout}
    (-15.75,-9) -- (11,-9)
    (-3.25,-9.5) node[text width=3cm, align = center]{RF / DC Schematic}
    
    ;
    \draw[-latex]
    (-8.25,-12.5) -- (-7.25,-12.5);

    \draw[-latex]
    (-7.25,-13.5) -- (-8.25,-13.5)
    ;

    \draw[dashed] 
    (-3,1) node[text width=1.7cm, align = center]{1K}
    (-11,1) node[text width=1.7cm, align = center]{4K}
    (6.75,1) node[text width=1.7cm, align = center]{4K}
    (9,1) node[text width=1.7cm, align = center]{300K}
    (-14,1) node[text width=1.7cm, align = center]{300K}
    (5.75,1) -- (5.75,-8.8)
    (-10.25,1) -- (-10.25,-8.8)
    (-11.75,1) -- (-11.75,-8.8)
    (7.75,1) -- (7.75,-8.8)
    ;
   \fill[black!20!blue, opacity=0.1] (-8.25,-3.5) rectangle (-5.25,-8.85);
   \fill[black!50!green, opacity=0.1] (-3.25,0) rectangle (3.5,-8.85);

\end{circuitikz}
\caption{Schematic of a single KPUP/SNSPD-pair with a representation on how the pair is read out digitally.  The green box encompassing the KPUP is shown in the schematic at the top of the image.  In the digital readout scheme, Tx and Rx stand for transceiver and receiver, and DAC and ADC stand for Digital-to-Analogue-Converter and Analogue-to-Digital-Converter, respectively.  For this demonstration, each KPUP and each SNSPD is biased separately.  For simplicity, additional filtering for both bias lines are omitted.  The SNSPD bias lines contain series resistors and capacitors to ground at different stages of the cryostat.  The KPUP bias lines have a series of low-pass LC filters before arriving at the chip.}
\label{sch}
\end{figure*}

Figure~\ref{sch} shows a schematic of the experimental setup for each SNSPD / KPUP pair.  The KPUP is a microstrip transmission line resonator that couples to an RF transmission line.  When a probe tone is sent through the RF chain within the cryostat, the KPUP will cause a dip in power at its resonant frequency.  The inductor in the circuit is a result of both the magnetic inductance and kinetic inductance.  When DC biased, as shown in Fig.~\ref{sch}, the kinetic inductance value increases, thereby decreasing the resonant frequency of the KPUP, according to $\omega_0 = 1/\sqrt{LC}$.  The kinetic inductance in a resonator that is made out of a non-linear material, such as NbTiN, will shift quadratically with bias current:

\begin{equation}
    L_K(I) = L_K(0) \bigg[1 + \frac{I_b^2}{I_2^2} + \frac{I_b^4}{I_4^4} ... \bigg],
\end{equation}

\noindent in which $I_b$ is the applied bias current and $I_2$ and $I_4$ are material and geometry depended parameters that set the scale of non-linearity of the film.  Following this equation, a kinetic inductance-based resonator becomes more responsive to additional bias as $I_b^2/I_2^2$ increases. \cite{kpup, Zmuidzinas-2012}

In Figure~\ref{sch}, each KPUP has an external current bias that is separate from the SNSPD bias.  The SNSPD is coupled to its KPUP pair via a 50$\Omega$ resistor.  When an additional signal is introduced to the KPUP, its resonant frequency will change according to:

\begin{equation}
    \omega (t) = \omega_c + \frac{d \omega}{d I_b} cos(\omega_s t),
\end{equation}

\noindent in which $\omega_c$ is the carrier or RF probe tone, $d\omega / d I_b$ is resonant frequency response from the KPUP bias, and $\omega_s$ is the external signal.  

When an SNSPD is coupled to the KPUP as the external signal and it has not been triggered, the resonant frequency is only a result of the current bias, $I_b$.  However, during a detection event the current that biases the SNSPD is re-routed from the SNSPD chip and through the inductor of the KPUP.  The resonant frequency shifts downward according to the increased current running through the KPUP before the SNSPD turns back to its superconducting state.

\subsection{Design and Fabrication}

The SNSPD chip used is a 1.5~$\mu$m wide wire with a 3~nm-thick sputtered WSi (W30Si70) film (30~s deposition) with a 3~nm aSi cap. Gold pads were applied using a liftoff method and then the micro-wires were defined using a deep-UV stepper. The wires were etched in a combination of CF4, CCl$_2$F$_2$ and O$_2$ before they were stripped of resist in acetone and then capped with another 3~nm of aSi.  The sheet resistance and sheet inductance were 1~k$\Omega$ and 520~pH/$\square$, respectively.  The chip has a 45\% fill factor on the wires and 250x1000~$\mu$m sized pixels.  A full characterization of similar microwires can be found in~\cite{Jaime2023}.

The KPUP was designed as an inverted microstrip $\lambda/2$ resonator capacitively coupled to a microwave feedline via the center of the transmission line resonator. The transmission line resonant mode has voltage nodes at the ends of the microstrip line and an antinode at the midpoint where the resonator is coupled to the feedline.  A band stop filter, designed to produce a low input impedance at the KPUP frequency, and then bond pads are connected to either end of the resonator.  The resonator circuit is made out of 35~nm-thick NbTiN, with 100~nm of amorphous silicon as the dielectric and 200~nm of Nb as the ground layer.  The KPUP resonator used in this demonstration is made of NbTiN with $I_c$ = 1.25~mA, $I_2$ = 3.67~mA, and $I_4$ = 2.99~mA.  This indicates that the maximum fractional change in kinetic inductance for this device is 5.8\%.

\subsection{Readout}

To read out one resonator in a typical analogue homodyne system, a signal generator produces an RF tone at or around the resonant frequency and gets split before going into the cryostat.  Half of the split tone goes through the RF lines within the cryostat before arriving at the RF port of an IQ mixer while the other split tone is routed directly to the LO port of the same IQ mixer.  The signal in the RF port is then mixed with the LO signal down to baseband and read out with an oscilloscope.  Any phase offset accumulated from going through the cryostat and resonator compared to the reference cable (LO) is shown in the I and Q timestreams after the IQ mixer.  Additional filtering and amplification may be applied to the I/Q lines before they arrive at the oscilloscope.

For frequency domain multiplexing, a digital readout system applies the same principles.  Fig.\ref{sch} shows the digital readout scheme used in this paper.  An RF tone is monitored by a GPU-based Software Defined Radio (SDR).  The Ettus Research X-300 SDR contains a Kintex-7 FPGA and additional RF daughterboards, both of which are responsible for signal generation, amplification, and data digitization.  All other operations, such as the remainder of the digital signal processing tasks, are performed by the host computer.  The available bandwidth for this experiment using the Ettus Research SDR is 100MHz.  The code for the digital readout was developed by Caltech~\cite{SDR} and has been used to readout Thermal Kinetic Inductance Detectors (TKIDs) and Quantum Capacitance Detectors (QCDs)~\cite{Minutolo2019}.

\section{Results}

\subsection{Frequency Domain Readout Demonstration}

\begin{figure}
     \centering
     \begin{subfigure}[b]{0.95\columnwidth}
         \centering
         \includegraphics[width=\columnwidth]{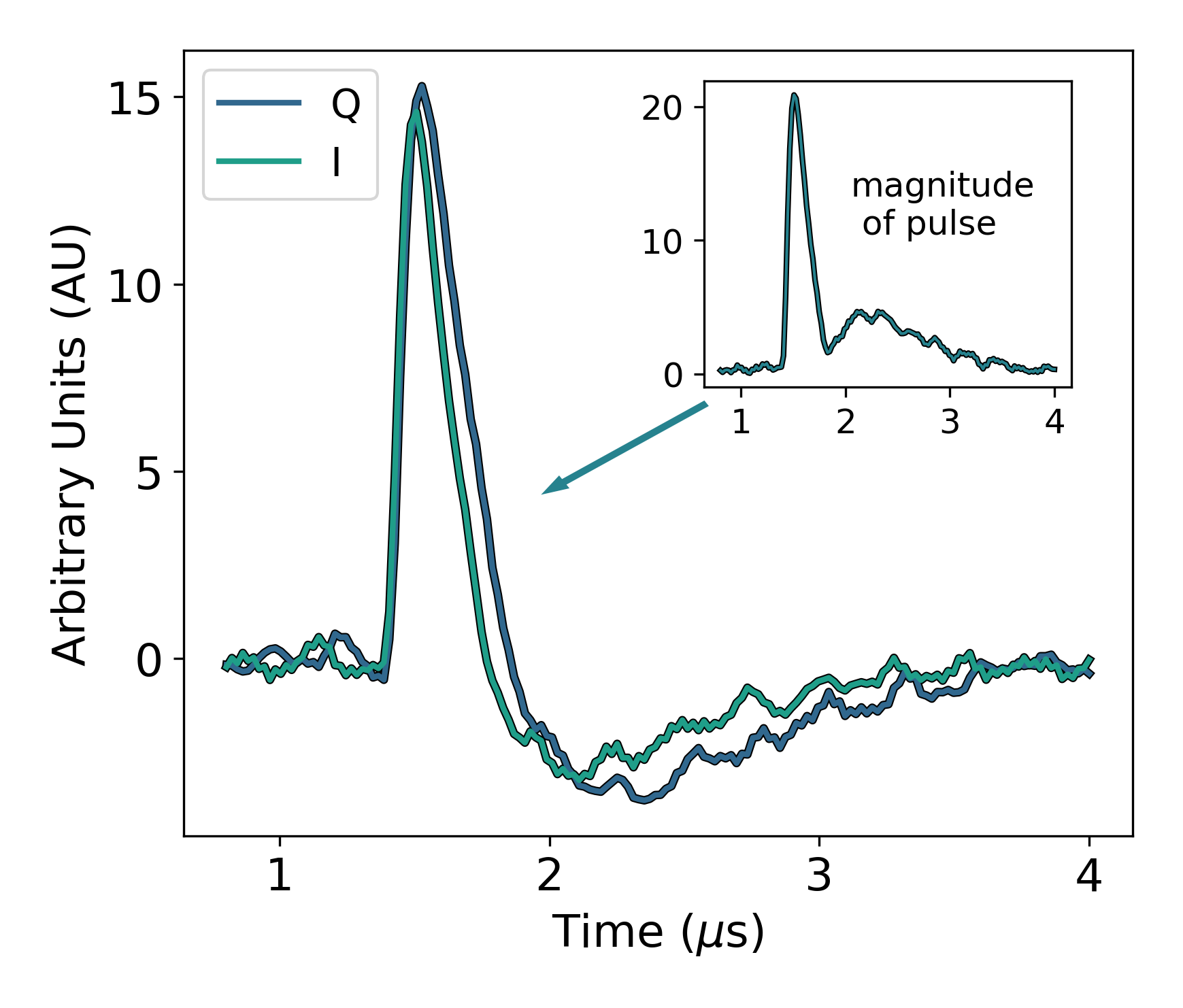}
         \label{fig:a}
     \end{subfigure}
     
     \begin{subfigure}[b]{0.95\columnwidth}
         \centering
         \includegraphics[width=\columnwidth]{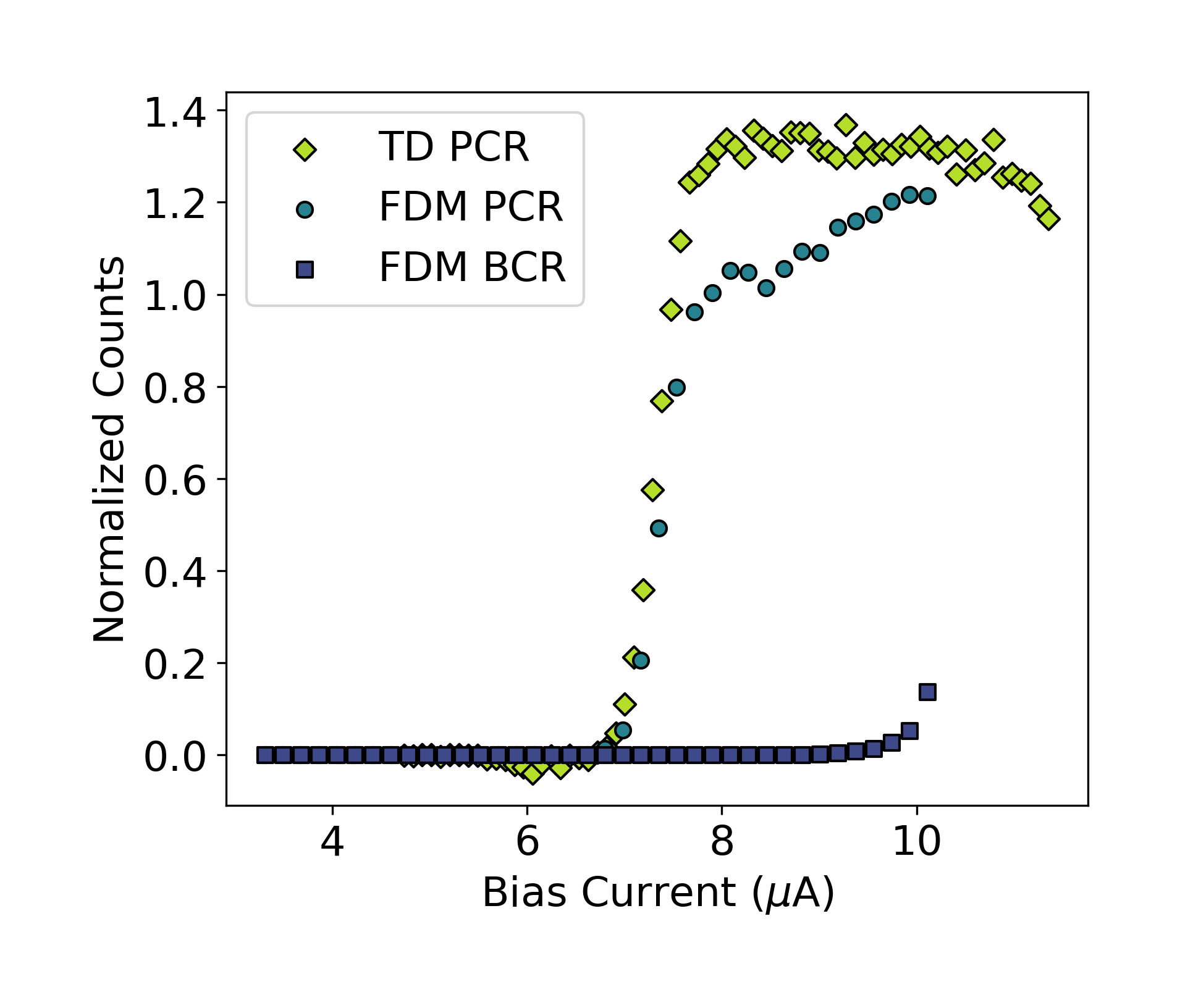}
         \label{fig:b}
     \end{subfigure}
        \caption{a) SNSPD pulse as seen by KPUP once demodulated to I and Q. b) Photon count rate (PCR) curve measured in time domain (TD) and by the KPUP (FDM), and background count rate (BCR) curve measured by the KPUP.}
        \label{pulse_comp}
\end{figure}

To monitor for pulses, an RF probe tone is sent from the SDR, through the RF chain containing the KPUP, and back to the SDR for down-conversion to baseband.  When there is no detection event, the I and Q values will be at some unchanging voltage level.  When a photon hits the SNSPD, I and Q will momentarily shift due to the change in amplitude and phase of the probe tone. 

Figure~\ref{pulse_comp} shows a pulse in the time domain using the KPUP in a digital readout architecture.  I and Q indicate the output timestream data from the digital IQ mixer in the SDR and the small pulse in the corner is the magnitude of I and Q.  The digital readout pulse was sampled at 100~MHz and digitally filtered with a bandpass of 1-10~MHz.  To calculate the rise time of the pulse, we find the time difference between 20\% and 80\% of the pulse height, which resulted in a 50.1~ns rise time.  The 1/e time on the recovery side of the pulse is 280~ns, which corresponds reasonably well to the theoretical value of 230~ns, based on the calculated 520~nH/$\square$ sheet inductance and 100~$\Omega$ of coupling resistance in between the KPUP and SNSPD.

The pulses are detected by using a simple threshold trigger on the magnitude of the pulses.  Any counts shorted than the length of a single pulse are disregarded.  Each pulse is time-tagged and the time-stream data around the pulse is saved for later reference.

Figure~\ref{pulse_comp} b) shows the photon count rate (PCR) and background count rate (BCR) of the SNSPD pixel using a conventional counter method (TD) and using the KPUP via the SDR (FDM). The SNSPD chip was illuminated with 470~nm light. The measurements were done on two separate cooldowns, and the operating base temperature for the FDM measurement was 930~mK, whereas the TD measurement was taken with an operating temperature of 800~mK.  The difference in operating temperature accounts for the difference in switching current between the two measurements.

\subsection{Parameter Space Investigation}

\begin{figure}
     \centering
     \begin{subfigure}[b]{0.95\columnwidth}
         \centering
         \includegraphics[width=\columnwidth]{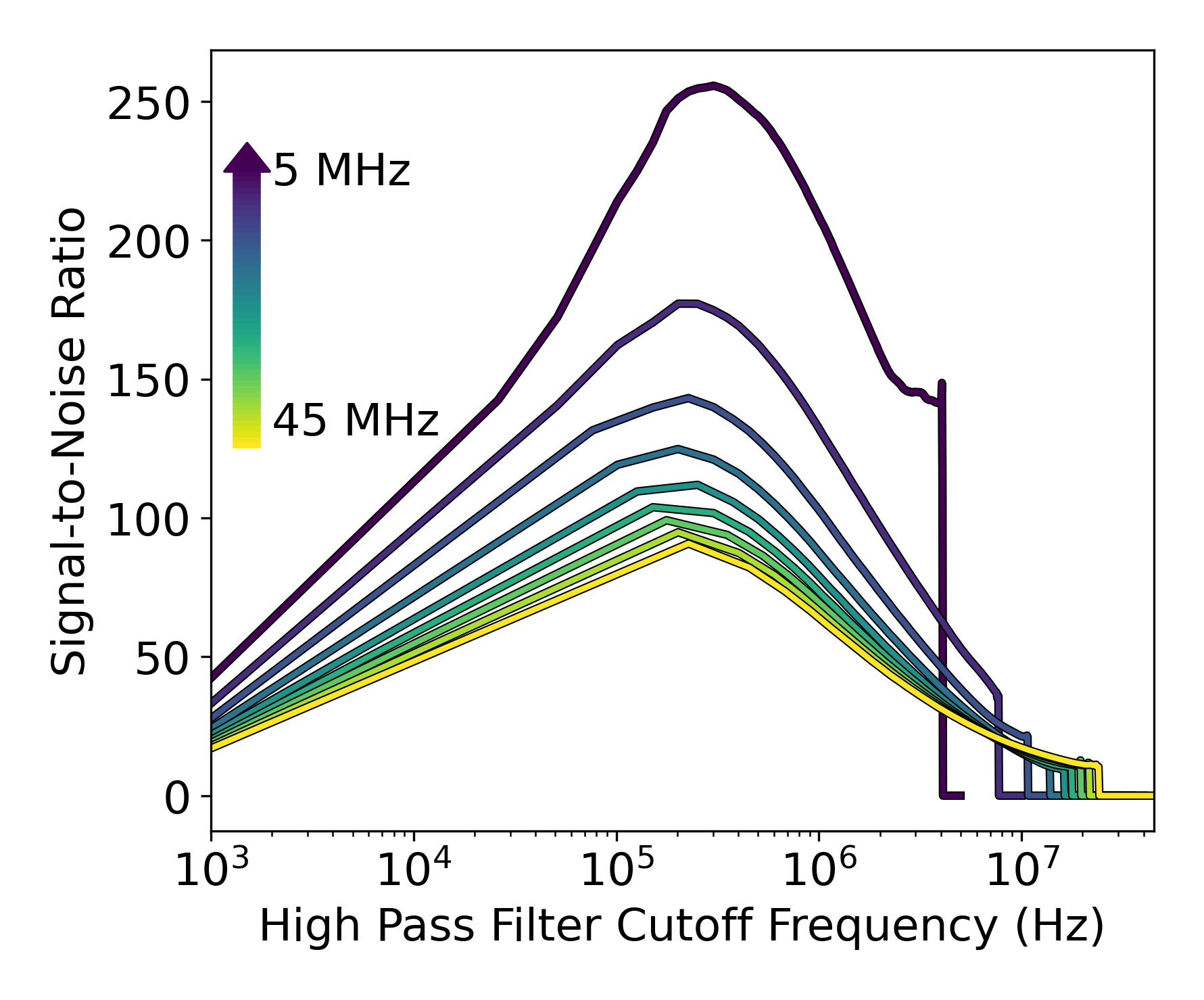}
     \end{subfigure}
     
     \begin{subfigure}[b]{0.95\columnwidth}
         \centering
         \includegraphics[width=\columnwidth]{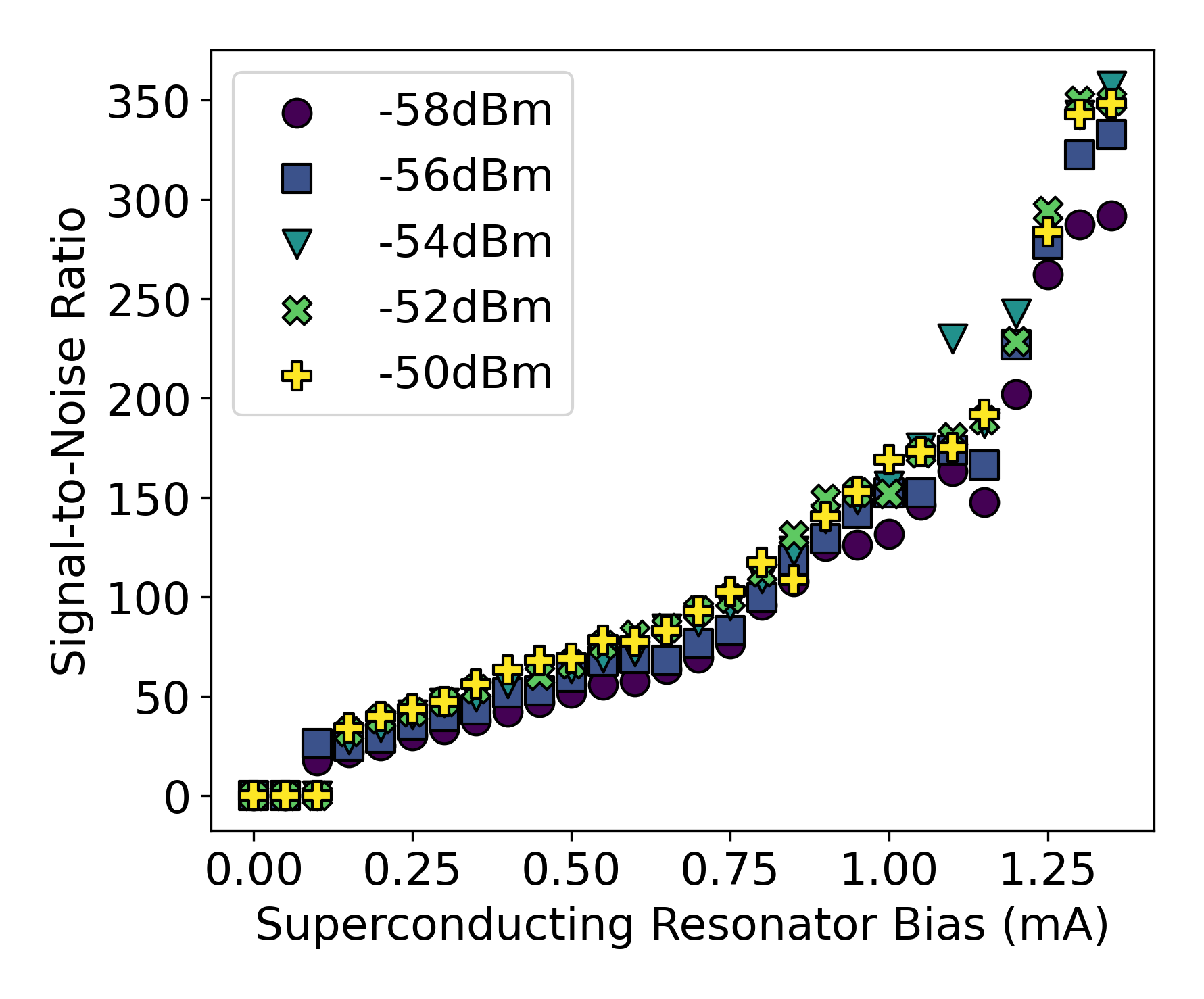}
     \end{subfigure}
        \caption{SNR of an SNSPD as a function of various parameters using a digital readout architecture. a) SNR of various filter cutoff frequencies using a 1-pole digital Butterworth filter.  The x-axis represents the high pass filter component and each line is a result of a different low pass filter cutoff frequency. b) SNR as a function of RF and DC bias into the KPUP resonator.}
        \label{param}
\end{figure}

To pave the road for multiplexed readout of SNSPDs with pulse heights on the $\mu A$ scale in the future, some parameters were changed to find the best signal-to-noise ratio (SNR).  The parameters changed include the RF and DC bias for the KPUP and the digital band pass filter cutoff frequencies during data acquisition.  Fig.~\ref{param} a) shows that for an SNSPD pulse of ~$10~\mu A$, the SNR can reach upwards of 250 when the bandpass filter is 300~kHz-5MHz.  Fig.~\ref{param} b) shows an obvious upward trend for SNR as both the RF input power and DC bias is increased.  The non-linear improvement in the SND can be explained by Eq. 1.

SNR was calculated by: 

\begin{equation}
    SNR = \frac{\bar{S}}{\sqrt{\sigma}},
\end{equation}

\noindent in which $\bar{S}$ is the mean value of the pulse amplitude and $\sigma$ is the standard deviation of the noise, considering that the length of the noise data set is equal to the number of points in the pulse.  To calculate SNR, we used a peak finder to find all of the pulses and then collected data sets of each pulse with 2$\mu$s of data before and after each pulse.  We used the data before the pulse as the noise data.  The data points that are considered the signal are any value that is higher than maximum value in the noise data set.  

\subsection{Multi-Tone Readout}

\begin{figure}
\includegraphics[width=1\columnwidth]{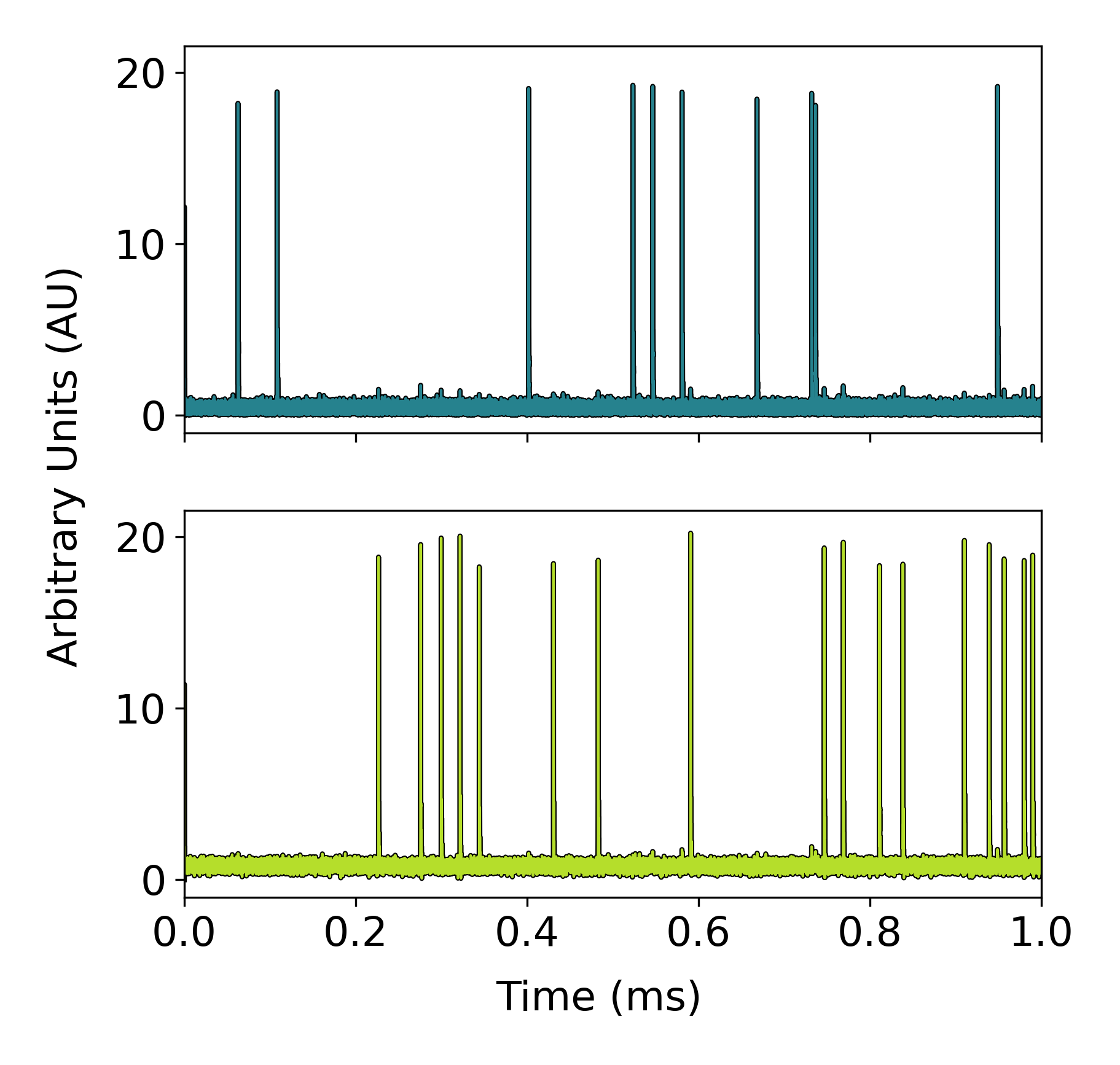}
\caption{Two-pixel demonstration of KPUP/SNSPD array using the SDR.}
\label{pixels}
\end{figure}

Figure~\ref{pixels} shows the amplitude response of two resonators and SNSPDs measured at the same time. For this measurement, we used a 16-channel DAC to differentially DC bias the SNSPDs and KPUPs.  Because both devices are biased with floating grounds, the Linduino can bias 4 SNSPDs and 4 KPUPs at the same time. 

\section{Future Work}

In this paper, we were only able to read out two pixels because of the limited bandwidth of the SDR (100 MHz instantaneous bandwidth).  FDM of single photon counting MKIDs has already been demonstrated using a previous generation readout architecture known as a Reconfigurable Open Architecture Computing Hardware (ROACH) board, with up to 2000 resonators on a single readout line~\cite{Fruitwala_2020}.  The next step is to increase the number of pixels multiplexed by switching to a Radio Frequency System on Chip (RFSoC) board, which is the latest generation of readout systems with instantaneous bandwidths of 2.5-5~GHz~\cite{Smith2022}. By allocating 10-25~MHz per resonator, up to 100-500 SNSPDs could be readout on a single feedline with such a system. The resonator bandwidth defines it's intrinsic ring-down time, which should be matched to the expected count rate of the SNSPD. The count rate per channel and total number of channels can be thus be traded off. Open-source firmware for superconducing resonator readout like this has already been largely developed for optical MKIDs and can thus be leveraged by such systems~\cite{Stefanazzi2022, qick}. Since the pulse amplitude of SNSPDs is linear with the supplied bias current, the SNR values of 350 that we achieved while operating the SNSPDs at 10~$\mu$A implies that the system will be capable of reading out sub-1~$\mu$A pulses. This makes the architecture ideally suited for use with mid- or far-IR SNSPD arrays in the future.

\begin{acknowledgments}
This research was performed at the Jet Propulsion Laboratory, California Institute of Technology, under contract with the National Aeronautics and Space Administration (NASA - 80NM0018D0004). S.S. research was supported by appointment to the Future Investigators in NASA Earth and Space Science and Technology (FINESST) program under contract with NASA. S.P. was supported by the National Science Foundation Graduate Research Fellowship. A.B.W. was supported in part by an appointment to the NASA Postdoctoral Program at the Jet Propulsion Laboratory, administered by Universities Space Research Association under contract with NASA. Support for this work was provided in part by the DARPA DSO Invisible Headlights and the NASA ROSES-APRA programs. The authors would like to thank Ioana Craiciu for technical assistance with, and maintenance of, a cryogenic probe station used for screening of the SNSPD chips. 
\end{acknowledgments}

\section{References}
\bibliography{references.bib}

\end{document}